\documentstyle[12pt]{article}

\begin{document}
\date{}

\title{On dynamical chiral symmetry breaking in quantum
electrodynamics}

\author{V.E. Rochev\\
{\it Institute for High Energy Physics}\\
{\it Protvino, Moscow Region, Russia}}

\maketitle

\newcommand{\be}{\begin{equation}}
\newcommand{\ee}{\end{equation}}
\newcommand{\ba}{\begin{eqnarray}}
\newcommand{\ea}{\end{eqnarray}}
\newcommand{\tr}{\,\mbox{tr}\,}

\begin{abstract}
The problem of dynamical chiral symmetry breaking (DCSB) in 
multidimensional quantum electrodynamics (QED) is considered.
It is shown that for six-dinesional QED the phenomenon of DSCB exists
in ladder model for any coupling.  
\end{abstract}

\newpage

\section*{Introduction}

Dynamical chiral symmetry breaking (DCSB) is an important topic of
particle physics. DCSB is a foundation of light hadron theory --
chiral dynamics, which is a low-energy limit of quantum
chromodynamics. It also plays an important role in generalizations of
Standard model.

One of the most studied examples of DCSB is the four-dimensional
massless
quantum electrodynamics (QED) in the strong coupling regime
\cite{Mas,
Fom}.
(A detailed consideration and extensive references can be found in
the monograph
\cite{Mir}.) 

In this paper an attempt is made to study the phenomenon of DCSB in
the six-dimensional QED. The interest in multidimensional model is
stimulated by rather popular investigations of theories with extra
dimensions (see \cite{Rub} for review). The mechanism of DCSB for
the six-dimensional QED, which is investigated in present paper, can
be
useful for understanding of multidimensional dynamics.

An investigation of nonperturbative effects, such as DCSB, in
the multidimensional QED assumes some model approximation. As  such
an
approximation we will use  a leading term of nonperurbative
expansion, elaborated in works \cite{Ro1, Ro2}. In the diagram
language the
leading order equation for electron propagator corresponds to
the well-known ladder approximation. Just in the framework of this
approximation  the phenomenon of DCSB have been investigated firstly
in the four-dimensional QED \cite{Mas, Fom}, and followed
investigations 
have achieved that the approximation quite adequately described this
effect. We suppose that in the six-dimensional QED such an
approximation
is also adequate to the situation.

The principal result of this paper is a conclusion about the
existence of
DCSB phenomenon for the six-dimensional QED. In contrast to
the four-dimensional QED, where a critical coupling constant  
$\alpha_c\sim 1$ exists (i.e., at $\alpha<\alpha_c$ the DCSB
phenomenon is absent), for the six-dimensional QED the phenomenon
exists
at any coupling. 

One of the principal problems, which arises in six-dimensional QED
studying, is a problem of renormalizability. In the framework of the
coupling
constant perturbation theory the six-dimensional QED is a
non-renormalizable theory. Nevertheless, this fact is not, generally
speaking, an obstacle for the existence of renormalized expansions of
another type. Thus, for example, renormalized
$1/N$-expansion exists for some models, which are non-renormalizable
in the usual sense of the coupling constant perturbation theory (see,
for example, \cite{Zinn}). We suppose that similar situation can be
realized also for gauge theories: nonperturbative expansions can be
sensible for multi-dimensional gauge theory, for which the usual
perturbative series does not exist.

\section{Schwinger-Dyson equations and iteration scheme}             

We  consider a theory of massless spinor field
$\psi(x)$ (electron) interacting with Abelian gauge field 
$A_\mu(x)$ (photon) in
 $D$-dimensional Minkowsky space with a metric  
$x^2\equiv x_\mu x_\mu =
x_0^2-x_1^2-\cdots-x_{D-1}^2$. (For simplification of notation all
vector indicies are written as lower ones) 
A Lagrangian (including a gauge fixing term) is

\be                                              
{\cal L} = -\frac{1}{4}F_{\mu\nu}F_{\mu\nu}
-\frac{1}{2d_l}(\partial_\mu A_\mu)^2
+ \bar\psi(i\hat\partial + e\hat A)\psi.
\label{lagrangian}
\ee 

Here $F_{\mu\nu}=\partial_\mu A_\nu - \partial_\nu A_\mu,\;
\hat A \equiv A_\mu\gamma_\mu;\; \bar\psi = \psi^*\gamma_0,\; 
e$ is coupling, $d_l$ is a gauge parameter,
 $\gamma_\mu$ are the Dirac matricies.

A generating functional of Green functions (vacuum expectation values
of $T$-products of fields) can be represented as the functional
integral
\be
G(J,\eta) = \int D(\psi,\bar\psi,A)
\exp i\Big\{\int dx\Big({\cal L}+J_\mu(x)
A_\mu(x)\Big)
-\int dx dy \bar\psi^\beta(y)\eta^{\beta\alpha}(y,x)
\psi^\alpha(x)\Big\}.
\label{G}
\ee

Here $J_\mu(x)$ is a gauge field source,
$\eta^{\beta\alpha}(y,x)$ is a bilocal source of spinor field
($\alpha$ and $\beta$ are spinor indicies). Normalization constant is
ommitted. 
Functional derivatives of $G$ over sources are vacuum expectation
values:
\be
\frac{\delta G}{\delta J_\mu(x)} = i<0\mid A_\mu(x)\mid 0>,\;\;
\frac{\delta G}{\delta\eta^{\beta\alpha}(y,x)} =
i<0\mid T\Big\{\psi^\alpha(x)\bar\psi^\beta(y)\Big\}\mid 0>.
\label{VEV}
\ee
Functional-derivative  Schwinger-Dyson equations (SDEs) for
generating functional $G$ read as follows
\be
(g_{\mu\nu}\partial^2-\partial_\mu\partial_\nu+
\frac{1}{d_l}\partial_\mu\partial_\nu)
\frac{1}{i}\frac{\delta G}{\delta J_\nu(x)} + 
ie\, \mbox{tr}\Big\{ \gamma_\mu\frac{\delta G}{\delta\eta(x,x)}\Big\}
+ J_\mu(x)G = 0,
\label{SDEA}
\ee
\be
\delta(x-y)G + i\hat\partial\frac{\delta G}{\delta\eta(y,x)}
+ \frac{e}{i}\gamma_\mu\frac{\delta^2G}{\delta J_\mu(x)
\delta\eta(y,x)} - \int dx' \eta(x,x')
\frac{\delta G}{\delta\eta(y,x')} = 0.
\label{SDEpsi}
\ee

Let resolve SDE (\ref{SDEA}) with respect to the first derivative of
generating functional over $J_\mu$:
  \be
  \frac{1}{i}\frac{\delta G}{\delta J_\mu(x)}=
  -\int dx_1 D^c_{\mu\nu}(x-x_1)\Big\{J_\nu(x_1)G+
  ie\tr \gamma_\nu\frac{\delta G}{\delta\eta(x_1,x_1)}\Big\}
  \label{dG/dJ}
  \ee
   and put the result into the second SDE (\ref{SDEpsi}).
   (Here
   $
   D^c_{\mu\nu} = [g_{\mu\nu}\partial^2-\partial_\mu\partial_\nu+
\frac{1}{d_l}\partial_\mu\partial_\nu]^{-1}
$ is a free photon propagator.)
  As a result we obtain "integrated over $A_\mu$"
  equation

\ba
\delta(x-y)G + i\hat\partial\frac{\delta G}{\delta\eta(y,x)}
+ \frac{e^2}{i}
\int dx_1 D^c_{\mu\nu}(x-x_1)\gamma_\mu 
\frac{\delta}{\delta\eta(y,x)}\tr \gamma_\nu\frac{\delta G}
{\delta\eta(x_1,x_1)}= 
\nonumber\\
= \int dx_1 \Big\{\eta(x,x_1)
\frac{\delta G}{\delta\eta(y,x_1)} + 
e D^c_{\mu\nu}(x-x_1)J_\nu(x_1)\gamma_\mu
\frac{\delta G}{\delta\eta(y,x)}\Big\}.
\label{SDE}
\ea 
Exploiting fermi-symmetry condition, one can rewrite  equation
(\ref{SDE}) in the form
 \ba
\delta(x-y)G + i\hat\partial\frac{\delta G}{\delta\eta(y,x)}
+ ie^2
\int dx_1 D^c_{\mu\nu}(x-x_1)\gamma_\mu 
\frac{\delta}{\delta\eta(x_1,x)}\gamma_\nu\frac{\delta G}
{\delta\eta(y,x_1)}= 
\nonumber\\
= \int dx_1 \Big\{\eta(x,x_1)
\frac{\delta G}{\delta\eta(y,x_1)} + 
e D^c_{\mu\nu}(x-x_1)J_\nu(x_1)\gamma_\mu
\frac{\delta G}{\delta\eta(y,x)}\Big\}.
\label{SDEF}
\ea 

For solution of SDE
(\ref{SDEF}) we shall use the iteration scheme  proposed in works
\cite{Ro1},
\cite{Ro2}.
A general idea of this scheme is an approximation of
functional-differential equation (\ref{SDEF}) by an equation with
"constant", i.e.
independent of the sources
$J_\mu$  and $\eta$, coefficients. Thus we approximate the
functional-differential SDE near the point
$J_\mu=0,\;\eta=0$. Since the objects of calculations are Green
functions,
i.e., derivatives of $G$ in zero, such an approximation seems to be
quite
natural.

In every order of this scheme the Green functions can be found as
solutions of a closed system of equations.

In correspondence with aforesaid, we choose as the leading
approximation to
equation (\ref{SDEF}) the following equation: 

 \be
\delta(x-y)G^{(0)} + i\hat\partial\frac{\delta G^{(0)}}
{\delta\eta(y,x)}
+ ie^2
\int dx_1 D^c_{\mu\nu}(x-x_1)\gamma_\mu 
\frac{\delta}{\delta\eta(x_1,x)}\gamma_\nu\frac{\delta G^{(0)}}
{\delta\eta(y,x_1)}= 0.
\label{G(0)F}
\ee 

A solution of this equation 
is the functional
\be
G^{(0)}=\exp\Big\{\mbox{Tr}(S\star\eta)\Big\}.
\label{G0F}
\ee
(The sign
$\star$ denotes a multiplication in operator sense, and the sign
$\mbox{Tr}$ denotes a trace in operator sense.)

Here
\be
S^{-1}(x)=-i\hat\partial\delta(x)
-ie^2D^c_{\mu\nu}(x)\gamma_\mu S(x)\gamma_\nu.
\label{charF}
\ee
 Equation (\ref{charF}) is an equation for elecron propagator in
 the leading approximation of the iteration scheme.

In correspondence with  (\ref{SDEF}) and
(\ref{G(0)F}) iteration equation  is
\ba
\delta(x-y)G^{(i)} + i\hat\partial\frac{\delta G^{(i)}}
{\delta\eta(y,x)}
+ ie^2
\int dx_1 D^c_{\mu\nu}(x-x_1)\gamma_\mu 
\frac{\delta}{\delta\eta(x_1,x)}\gamma_\nu\frac{\delta G^{(i)}}
{\delta\eta(y,x_1)}= 
\nonumber\\
= \int dx_1 \Big\{\eta(x,x_1)
\frac{\delta G^{(i-1)}}{\delta\eta(y,x_1)} + 
e D^c_{\mu\nu}(x-x_1)J_\nu(x_1)\gamma_\mu
\frac{\delta G^{(i-1)}}{\delta\eta(y,x)}\Big\}.
\label{G(i)F}
\ea 
 Equation (\ref{charF}) and  equations for the higher Green
 functions,
 which followed from (\ref{G(i)F}), in diagram language 
 correspond to the well-known ladder approximation. In our treatment
 these equations are a consistent part of the iteration scheme.

\section{Asymptotic solution of electron propagator equation and
dynamical chiral symmetry breaking}

In transverse Landau gauge 
$d_l=0$ electron propagator equation (\ref{charF}) has a simple
solution
\be
 S_0=-1/i\hat\partial.
\label{Sc0}
\ee
Really, in coordinate space
\be
D^c_{\mu\nu}(x)=\frac{e^{-i\pi D/2}\Gamma(D/2-1)}
{4i\pi^{D/2}(x^2-i0)^{D/2-1}}\Big[\frac{1+d_l}{2}g_{\mu\nu}
+(1-d_l)\big(D/2-1\big)\frac{x_\mu x_\nu}{x^2-i0}\Big].
\label{Dcx}
\ee
At  $d_l=0$ the function $D^c_{\mu\nu}(x)$
possesses an important property ("$\hat x$-transversality") 
\be
D^c_{\mu\nu}(x)\gamma_\mu\hat x\gamma_\nu = 0,
\label{Dcxprop}
\ee
from which the existence of solution (\ref{Sc0}) follows immediately,
since 
$S_0(x)\sim \hat x$. 

At $D$ even one can define chiral components of spinor fields and
corresponding chiral transformations. Lagrangian (\ref{lagrangian})
of massless electrodynamics is invariant in respect of the chiral
transformations. Solution (\ref{Sc0}) is the chiral-symmetric one.
The existence of non-chiral-symmetric solutions of equation
(\ref{charF}) with $\tr S\neq 0$ denotes dynamical chiral symmetry
breaking (DCSB) in this model.

Equation (\ref{charF}) is a nonlinear equation. This fact 
leads to essential difficulties for its investigation. However,
an expierence of investigation of this equation at $D=4$ (see
\cite{Mir}
and refs. therein) demonstrates that an ultraviolet behaviour of its
solution is
defined by the linear approximation. Since for QED the
nonperturbative
region is the ultraviolet region, then a linearized version of this
equation is quite enough  for description of
nonperturbative effects (such as DCSB) . At $D=4$ this question was
investigated in
detail (see \cite{Mir}). We accept this supposition also for $D>4$
and shall investigate a linerization which is known as the
bifurcation 
approximation \cite{Gus}.
 
Let introduce mass operator
$$
\Sigma = S^{-1}-S^{-1}_0.
$$
The linearization prosedure consists in the following approximation
$$
S=[S^{-1}_0+\Sigma]^{-1}\approx S_0 -
S_0\star\Sigma\star S_0.
$$
For the mass operator in transverse gauge we obtain equation
\be
\Sigma=ie^2D_{\mu\nu}\gamma_\mu (S_0\star\Sigma\star S_0)\gamma_\nu.
\label{Sigma}
\ee

Due to construction above it is evident  that a region of
applcability
of
this linerized version is an asymptotic ultraviolet region, i.e.
the momentum
region $p^2\gg\lambda^2$, where $\lambda$ is a mass parameter which
plays a role of infrared cutoff. As a particular consequence one gets
that in the region of applicability the solutions of equation
(\ref{Sigma}) should fulfil the condition  
$$
\Sigma^2\le p^2.
$$
Due to the condition of $\hat x$-transversality (\ref{Dcxprop}) a
spinor
structure of solution of equation (\ref{Sigma}) is trivial:
$$
\Sigma_{\alpha\beta}=I_{\alpha\beta}\cdot\Sigma,
$$
and therefore
$$
F_{\alpha\beta}\equiv (S_0\star\Sigma\star S_0)_{\alpha\beta}
=I_{\alpha\beta}\cdot F,
$$
and finally we obtain for the mass operator  the following equation
in
$x$-space:
$$
\Sigma(x^2)=\alpha\frac{(D-1)e^{-i\pi D/2}\Gamma(D/2-1)}
{[\pi(x^2-i0)]^{D/2-1}}\cdot F(x^2).
$$
Here $\alpha=e^2/4\pi.$

Multiplaying 
\footnote{Such a multiplication is, in essence, some regularization
of  singular product $(x^2-i0)^{1-D/2}\cdot F(x)$. }
this equation by $(x^2)^{D/2-1}$ and passing to $p$-space, we obtain
the differential equation 
$$
(\partial^2)^{D/2-1}\Sigma(p^2)=-\alpha
\frac{(D-1)\Gamma(D/2-1)}{\pi^{D/2-1}(p^2+i0)}\cdot
\Sigma(p^2).
$$
This is an equation for the mass operator in the pseudo-Euclidean
Minkowsky
space. Performing Euclidean rotation
$\partial^2\rightarrow-\partial^2,\;p^2\rightarrow-p^2$,
we obtain the following equation for the mass operator in the
Euclidean momentum
space:
\be
(-\partial^2)^k\Sigma(p^2)=\alpha\frac{(2k+1)\Gamma(k)}
{\pi^kp^2}\cdot\Sigma(p^2),
\label{Sigmaeuc}
\ee
where $k=D/2-1$.

Consider firstly the four-dimensional case ($k=1$). In this case at
$\alpha<\pi/3$
equation (\ref{Sigmaeuc})  has  asymptotic solution
$$
\Sigma=C(p^2)^a,
$$
where
\be
a=-\frac{1}{2}+\frac{1}{2}\sqrt{1-3\alpha/\pi}.
\label{degree}
\ee
At $\alpha\ge\pi/3$ a solution of the equation is
\be
\Sigma=\frac{C}{\sqrt{p^2}}
\sin\Bigl(\frac{\omega}{2}\log\frac{p^2}{M^2}\Bigr),
\label{Sigma4C}
\ee
where 
$\omega=\sqrt{3\alpha/\pi-1}$.
Here $C$ and $M$ are real-valued constants.  
(At  $\omega\rightarrow 0$
the solution is
$
\Sigma=\frac{C}{\sqrt{p^2}}
\log\frac{p^2}{M^2}.)
$

At critical point $\alpha_c=\pi/3$ the type of the asymptotics
changes -- it becames oscillating. This change of behaviour means a
phase transition to the state with dynamically broken chiral symmetry
(see \cite{Mir}). By ohter words, at  $\alpha<\alpha_c$ only trivial
solution
$\Sigma\equiv 0$ exists, and at $\alpha\ge\alpha_c$ a non-trivial
solution arises, which corresponds to DCSB. To illustrate this thesis 
consider a procedure of normalization of the solution in the
pseudo-Euclidean Minkowski space. In the pseudo-Eucidean space a mass
operator should satisfy the normalization condition
\be
\Sigma(m^2)=m.
\label{Nor}
\ee
An analytical continuation into the pseudo-Euclidean space
$p^2\rightarrow-p^2-i0$ is performed with well-known formulae 
\be
(-p^2-i0)^a=e^{-i\pi a}(p^2+i0)^a,\;\;
\log(-p^2-i0)=\log(p^2+i0)-i\pi.
\label{Anal}
\ee
Taking into account these formulae it is easy to see that for
solutions at $\alpha<\pi/3$ the normalization condition contradicts
to reality condition for $C$ and $m$ (at non-zero values of these
quatities). At the same time in region $\alpha\ge\pi/3$ a normalized
solution with $C\neq 0$ exists, which corresponds to DCSB. It has the
form (in the Euclidean space):
\be
\Sigma(p^2)=\frac{m^2}{\sinh(\frac{\pi\omega}{2})\sqrt{p^2}}
\sin\Bigl(\frac{\omega}{2}\log\frac{p^2}{m^2}\Bigr).
\label{Sigma4}
\ee

Let turn to the six-dimensional case. Equation (\ref{Sigmaeuc}) at
$k=2$
can be
rewritten as the Mejer equation 
\cite{Mejer}
\be
\Bigl(z\frac{d}{dz}+2\Bigr)\Bigl(z\frac{d}{dz}+1\Bigr)
\Bigl(z\frac{d}{dz}-1\Bigr)
z\frac{d\Sigma}{dz}-z\Sigma=0,
\label{Sigmadif6}
\ee
where $z=\frac{5\alpha}{(4\pi)^2}p^2.$ A real-valued fundamental
system of
solutions near the infinite point $z=\infty$ has the form
$$
u_1(z)=G^{40}_{04}(ze^{-4\pi i}\mid -2,-1,0,1),\;
$$
$$
u_2(z)=G^{40}_{04}(z\mid -2,-1,0,1),\;
$$
$$
u_{3,4}(z)=e^{i\phi}G^{40}_{04}(ze^{-2\pi i}\mid -2,-1,0,1)
+e^{-i\phi}G^{40}_{04}(ze^{2\pi i}\mid -2,-1,0,1).
$$
Here $G^{40}_{04}$ is the Mejer function,  $\phi$ is a real number.
Asymptotics of functions $u_l$ at $z\rightarrow\infty$ are 
$$
u_1(z)\sim z^{-7/8}\exp(4z^{1/4}),\;
$$
$$
u_2(z)\sim z^{-7/8}\exp(-4z^{1/4}),\;
$$
$$
u_{3,4}(z)\sim z^{-7/8}\cos(4z^{1/4}+\phi).
$$
Since equation (\ref{Sigmaeuc}) itself has the asymptotical
character,
we can consider these asymptotics as solutions of our problem. The
exponentially rising solution is not satisfy to the condition
$\Sigma^2\le p^2$ and should be ignored. Hence, the leading
asymptotic
solution is
$$
\Sigma\approx Cz^{-7/8}\cos(4z^{1/4}+\phi).
$$
On taking into account the analytic continuation formulae
(\ref{Anal}) normalization condition (\ref{Nor}) in the
pseudo-Euclidean space and
the reality condition for the constants $C$ and $\phi$ give us 
equations which connect  $C$ and $\phi$ with the mass $m$. Resolving
these equations we obtain in the Euclidean space the following
normalized
asymptotic solution:  
\be
\Sigma(p^2)\approx 2m\Bigl(\frac{m^2}{p^2}\Bigr)^{7/8}
\exp\Bigl\{(5\alpha)^{1/4}\sqrt{\frac{2m}{\pi}}\Bigr\}
\cos\Bigl\{\frac{2}{\sqrt{\pi}}(5\alpha)^{1/4}\Bigl(\sqrt{\vert
p\vert}
-\sqrt{\frac{m}{2}}\Bigr)-\frac{7}{8}\pi\Bigr\}.
\label{Sigma6}
\ee
Here $\vert p\vert\equiv\sqrt{p^2}.$

\section{Dynamical chiral symmetry breaking in ultraviolet cutoff
scheme}

The construction given above is based on asymptotic solutions of
differential equation (\ref{Sigmaeuc}) and normalization condition
(\ref{Nor}) and is, in essence, nothing else than heuristic
consideration.

For more complete motivation of our principal position on existence
of
DCSB
phase in the multidimensional electrodynamics we use
 the  general Bogoliubov method for elaborating
models with spontaneous symmetry breaking. In accordance with the
method we
shall consider the
problem with explicit breakdown of chiral symmetry. For this purpose
we introduce a
mass term $m_0\bar\psi\psi$ with "seed" mass $m_0$ into Lagrangian
(\ref{lagrangian}), and, after solution of the corresponding
asymptotic boundary problem, go to  chiral limit, i.e. tend $m_0$ to
zero. DCSB criterion will
be non-zero value of mass operator in such a chiral limit:

$$
\lim_{m_0\to 0}\Sigma\neq 0.
$$
Introducing of the seed mass $m_0$ results in modification of
the inhomogeneous term in equation
(\ref{Sigma}): $-i\hat\partial\rightarrow (m_0-i\hat\partial)$,
but does not change differential equation (\ref{Sigmaeuc}), since the
inhomogeneous term disappears after the multiplication by
$(x^2)^{D/2-1}$. The role of the seed mass consists in a modification
of
boundary conditions. To derive and take into account these boundary
conditions it is necessary to turn to an integral equation in the
momentum
space. The integral equation for mass operator in the Euclidean
momentum
space for the linearized version of the model under consideration has
the form:
\be
\Sigma(p^2)=m_0+e^2\frac{D-1}{(2\pi)^D}\int d^Dq
\frac{\Sigma(q^2)}{q^2}\frac{1}{(p-q)^2}.
\label{SigmaD}
\ee
Here $\Sigma$ is the renormalized mass operator, $e^2$ is the
renormalized
coupling. The seed mass $m_0$ is a function of regularization
parameter. 
In the definition of this mass a wave function renormalization
constant
and a counterterm of mass renormalization are included (see
\cite{Ro2}
for more detail).

To integrate over angles we use formula
$$
J_D\equiv\int d^Dq\frac{f(q^2)}{(p-q)^2}=
\frac{\pi^{\frac{D-1}{2}}}{\Gamma(\frac{D-1}{2})}
\int dq^2(q^2)^{D/2-1}f(q^2)\int_{0}^{\pi} d\theta
\frac{\sin^{D-2}\theta}{p^2+q^2-2\vert p\vert\vert q\vert\cos\theta}.
$$
At $D=4$:
$$
J_4=\pi^2\int dq^2 q^2 f(q^2)\Biggl(\frac{1}{p^2}\theta(p^2-q^2)
+\frac{1}{q^2}\theta(q^2-p^2)\Biggr),
$$
and at $D=6$:
$$
J_6=\frac{\pi^3}{6}\int dq^2(q^2)^2 f(q^2)
\Biggl(\frac{1}{p^2}\biggl(3-\frac{q^2}{p^2}\biggr)\theta(p^2-q^2)
+\frac{1}{q^2}\biggl(3-\frac{p^2}{q^2}\biggr)\theta(q^2-p^2)\Biggr).
$$

Consider  firstly the four-dimensional case. In a scheme with
ultraviolet cutoff the integral equation for the mass operator at
$D=4$
has the form
\be
\Sigma(p^2)=m_0+\frac{3\alpha}{4\pi}\int^{\Lambda^2} dq^2\Sigma(q^2)
\Biggl(\frac{1}{p^2}\theta(p^2-q^2)
+\frac{1}{q^2}\theta(q^2-p^2)\Biggr).
\label{Sigmaint4}
\ee
 This equation leads to a boundary condition at $p^2=\Lambda^2$
 (which
we shall name as the ultraviolet condition): 
\be
\frac{d}{dp^2}\biggl(p^2\Sigma(p^2)\biggr)\Bigg\vert_{p^2=\Lambda^2}=
m_0
\label{uv4}
\ee
Another boundary condition (at small $p^2$) does not contain the
parameter
$m_0$ and does not play any part in our construction. Integral
equation (\ref{Sigmaint4}) is reduced to the differential equation
\be
\frac{d^2}{d(p^2)^2}
\biggl(p^2\Sigma(p^2)\biggr)=
-\frac{3\alpha}{4\pi}\frac{\Sigma(p^2)}{p^2}.
\label{Sigmadif4}
\ee
It is easy to see that this equation is the same as equation
(\ref{Sigmaeuc}) at $k=1\;(D=4)$.

Consider a pre-critical case $\alpha<\pi/3$. In this case a general
solution of equation (\ref{Sigmadif4}) is 
$$
\Sigma=C_1(p^2)^a+C_2(p^2)^{-a-1},
$$
Here  $-1/2<a<0$ (see eq.(\ref{degree})).

Suppose firstly $C_1\neq 0$. Then from condition (\ref{uv4}) we see,
 to make the solution  independent of the cutoff parameter
$\Lambda$ it is necessary to  renormalize the seed mass 
\be
m_0(\Lambda)=\mu\Bigl(\frac{\Lambda^2}{\mu^2}\Bigr)^a,
\label{m0}
\ee
which gives the value $C_1=\frac{\mu^{1-2a}}{a+1}$. From 
(\ref{m0}) it follows that at the chiral limit we have
$\mu\rightarrow
0$, and, consequently, $C_1=0$. If $C_1=0$, the renormalization of
the seed mass is produced with the formula
$$
m_0(\Lambda)=\mu\Bigl(\frac{\Lambda^2}{\mu^2}\Bigr)^{-a-1}
$$
and condition (\ref{uv4}) give us $C_2=-\mu^{3+2a}/a$, and again in
the chiral limit we have $\mu\rightarrow 0$, and, consequently,
$C_2=0$.
So,  taking into account ultraviolet boundary condition
(\ref{uv4}) results in absence of nontrivial solutions in the chiral
limit.

In the critical region $\alpha\ge\pi/3$ a general solution of
equation
(\ref{Sigmadif4}) is given by formula (\ref{Sigma4C}). Boundary
condition (\ref{uv4}) in this case results in the following formula
of
the mass renormalization:
$$
m_0(\Lambda)= \frac{\mu^2}{2\Lambda}\Biggl(\sin\biggl(
\frac{\omega}{2}\log\frac{\Lambda^2}{M^2}\biggr)
+\omega\cos\biggl(\frac{\omega}{2}\log\frac{\Lambda^2}{M^2}\biggr)
\Biggr).
$$
If the following condition fulfils
\be
\tan\biggl(\frac{\omega}{2}\log\frac{\Lambda^2}{M^2}\biggr)
=-\omega,
\label{tg4}
\ee
then at the  chiral limit a nontrivial solution exists:
$$
\Sigma=\frac{\mu^2}{\sqrt{p^2}}
\sin\biggl(\frac{\omega}{2}\log\frac{p^2}{M^2}\biggr),
$$
which corresponds to DCSB phase.
After normalization of this solution in the pseudo-Euclidean space on
the physical mass $m$ we come back to normalized solution
(\ref{Sigma4}).

We see, that oscillating
character of solution in critical region $\alpha>\pi/3$ is the major
property 
 ensuring the existence of DCSB.

For the six-dimensional space the integral equation for the mass
operator has
the form
\be
\Sigma(p^2)=m_0+\frac{5\alpha}{6(4\pi)^2}\int^{\Lambda^2}
dq^2\;q^2\Sigma(q^2)
\Biggl(\frac{1}{p^2}\biggl(3-\frac{q^2}{p^2}\biggr)\theta(p^2-q^2)
+\frac{1}{q^2}\biggl(3-\frac{p^2}{q^2}\biggr)\theta(q^2-p^2)\Biggr).
\label{Sigmaint6}
\ee
Ultraviolet boundary conditions, which follow from this integral
equation, are 
\be
\frac{d^2}{d(p^2)^2}\biggl((p^2)^2
\Sigma(p^2)\biggr)\Bigg\vert_{p^2=\Lambda^2}=
2m_0,
\label{uv6-1}
\ee
\be
\frac{d^3}{d(p^2)^3}\biggl((p^2)^2
\Sigma(p^2)\biggr)\Bigg\vert_{p^2=\Lambda^2}=
0.
\label{uv6-2}
\ee
A differential equation, which follows from 
(\ref{Sigmaint6}), coincides with equation  (\ref{Sigmadif6}). On
taking into account condition $\Sigma^2\le p^2$, its solution is 
$$
\Sigma\approx (p^2)^{-7/8}\bigl(C_1\cos(\kappa
\sqrt{\vert p\vert}+\phi)+C_2\exp(-\kappa
\sqrt{\vert p\vert})\bigr),
$$
where $\kappa=2(5\alpha/\pi^2)^{1/4}$.

Boundary conditions (\ref{uv6-1}) and (\ref{uv6-2}) give relations
\be
C_2=C_1e^{\kappa\sqrt{\Lambda}}\sin(\kappa\sqrt{\Lambda}+\phi)
\label{C2}
\ee
and
$$
\Bigl(\frac{\kappa}{4}\Bigr)^2\Lambda^{-3/4}C_1
\bigl(\sin(\kappa\sqrt{\Lambda}+\phi)-
\cos(\kappa\sqrt{\Lambda}+\phi)\bigr)=2m_0.
$$
Consequently, the mass renormalization is made with the formula
$$
m_0=\mu\Bigl(\frac{\mu}{\Lambda}\Bigr)^{3/4}
\bigl(\sin(\kappa\sqrt{\Lambda}+\phi)-
\cos(\kappa\sqrt{\Lambda}+\phi)\bigr),
$$
and, under condition
\be
\tan(\kappa\sqrt{\Lambda}+\phi)=1,
\label{tg6}
\ee
as well as for the critical region of four-dimensional theory, at the
chiral limit $m_0=0$ a
nontrivial solution exists, which corresponds to
DCSB.

An analytic continuation into the pseudo-Euclidean region and
normalization condition (\ref{Nor}) give relations, which connect
the constants $C_1,\;C_2$ and $\phi$ with the physical mass  $m$.
These
relations are
$$
\cases{\frac{1}{2}C_1e^{\kappa\sqrt{\frac{m}{2}}}
\cos\Bigl(\kappa\sqrt{\frac{m}{2}}+\phi\Bigr)+
C_2e^{-\kappa\sqrt{\frac{m}{2}}}\cos\Bigl(
\kappa\sqrt{\frac{m}{2}}\Bigr)=m^{11/4}\cos\frac{7\pi}{8},
\cr
\frac{1}{2}C_1e^{\kappa\sqrt{\frac{m}{2}}}
\sin\Bigl(\kappa\sqrt{\frac{m}{2}}+\phi\Bigr)+
C_2e^{-\kappa\sqrt{\frac{m}{2}}}\sin\Bigl(
\kappa\sqrt{\frac{m}{2}}\Bigr)=-m^{11/4}\sin\frac{7\pi}{8}.
\cr}
$$
Exploiting condition (\ref{tg6}) and formula (\ref{C2}) in the region
of applicability of our constructions one can
prove the
following inequality:
$C_2e^{-\kappa\sqrt{\frac{m}{2}}}\ll 
C_1e^{\kappa\sqrt{\frac{m}{2}}}$.
Really,  taking into account equations (\ref{tg6}) and
(\ref{C2}), one can exclude the coefficients $C_1$
and $C_2$ from the above relations  and obtains the following
equation on the phase factor:
\be
\sin x = -\sqrt{2}\sin x_0\;e^{-x_0-x}
\label{phas}
\ee
Here the following notations are introduced:
$$
x_0=\kappa\sqrt{\frac{m}{2}}-\frac{\pi}{8},
$$
$$
x=\phi+x_0+\pi l,
$$
where $l$ is an entire number, which is produced by a solution of
condition (\ref{tg6}): $\kappa\sqrt{\Lambda}+\phi=\pi/4+\pi l$. 
In the asymptotic region the solution of equation (\ref{phas})
is
$$
x\approx\pi n,
$$
where $n$ is an entire  {\it positive} number. Taking into
account (\ref{tg6}) and (\ref{C2}), we obtain
$$
C_2e^{-\kappa\sqrt{\frac{m}{2}}}\approx
\frac{C_1}{\sqrt{2}}e^{\pi/8-\pi n}\cos \pi l\ll
C_1e^{\kappa\sqrt{\frac{m}{2}}}.
$$
Neglecting in correspondence with proven inequality the terms with
$C_2$ in the normalization condition, we go again to formula
(\ref{Sigma6}) for the mass operator in the Euclidean space.

\section*{Conclusion}

The principal result of this paper is a nonperturbative model
motivation for the existence of DCSB phenomenon in the
six-dimensional
QED.
This result is in want of further specification and
investigation. Thus, for example, it is not quite clear what is
happen at a cutoff removing for the six-dimensionsal case. In
the four-dimensional ladder QED, in correspondence with results of
investigations summarized in monograph  \cite{Mir} we have  
$$\alpha\rightarrow\alpha_c$$
at the
cutoff removing in the critical region, i.e.
the renormalized QED in the strong coupling regime exists at the
critical
coupling only. If one will proceed by analogy with
the four-dimensional theory in the six-dimensional case, it may be
expected, that
$\alpha\rightarrow 0$ at the cutoff removing (though DCSB phenomenon
is left, i.e. an electron get a mass). In this connection inavoidable
question about trviality arises (as in the four-dimensional theory,
though). A solution of these problems requires an investigation of
the Bethe-Salpeter equation for bound states, i.e., in terms of our
expansion, an investigation of equations of the following iteration
step.  

In conclusion let us touch on  the multidimensional QED with a
dimension greater than six. Equation (\ref{Sigmaeuc}) has the
oscillating asymptotic solution at any even $D\ge 6$. This fact gives
rise  supposition that  DCSB phenomenon exists at any even
dimension greater than  four.

Author is grateful to G.G. Volkov for stimulating discussion and P.A.
Saponov for reading the manucsript.

\end{document}